\documentstyle{article}
\date{}             
\begin{document}
\title{The $K^\ast K\pi$ and $\rho\pi\pi$ couplings in QCD}
\author{Shi-Lin Zhu\\
Institute of Theoretical Physics, Academia Sinica\\ 
P.O.BOX 2735, Beijing 100080, P.R.China}
\maketitle
\begin{center}
\begin{minipage}{120mm}
\vskip 0.6in
\begin{center}{\bf Abstract}\end{center}
{\large
The light cone QCD sum rules are derived for 
the $K^\ast K\pi$ coupling $g_{K^\ast K\pi} $ and 
the $\rho\pi\pi$ coupling $g_{\rho\pi\pi}$.
The contribution from the excited states and the continuum is subtracted 
cleanly through the double Borel transform with respect to the two external 
momenta, $p_1^2$, $p_2^2=(p-q)^2$. 
Our result $g_{K^\ast K\pi}=(8.7\pm 0.5)$ and 
$g_{\rho\pi\pi}=(11.5\pm 0.8)$ is in good 
agreement with the experimental value.

\vskip 0.5 true cm
PACS: 13.75.Lb---Meson-meson interactions; 14.40.-n---Mesons; 
13.25.-k---Hadronic decays of mesons; 12.38.Lg---Nonperturbative methods
}
\end{minipage}
\end{center}
\large
\section{Introduction}
\label{sec1}
The $\rho\pi\pi$ ($K^\ast K\pi$) coupling $g_{\rho\pi\pi}$ ($g_{K^\ast K\pi}$)
plays a very important role in the phenomenological models for 
the nuclear force and nuclear matter. Now it is widely believed that 
QCD is the underlying theory of the strong interaction.
Yet the complicated infrared behavior of QCD causes the first principle 
derivation of hadron properties highly nontrivial. 
So a quantitative calculation of the $g_{\rho\pi\pi}$ ($g_{K^\ast K\pi}$)
coupling with a tractable and reliable theoretical approach proves valuable.

The method of QCD sum rules (QSR), as proposed
originally by Shifman, Vainshtein, and Zakharov \cite{SVZ} and adopted,
or extended, by many others \cite{RRY,IOFFE,BALIT}, are very 
useful in extracting the low-lying hadron masses and couplings.
In the QCD sum rule approach the nonperturbative QCD effects 
are partly taken into account through various condensates 
in the nontrivial QCD vacuum. In this work we shall use the light cone 
QCD sum rules (LCQSR) to calculate the $K^\ast K\pi$ and $\rho\pi\pi$ couplings.

The LCQSR is based on the OPE on the light cone, 
which is the expansion over the twists of the operators. The main contribution
comes from the lowest twist operator. Matrix elements of nonlocal operators 
sandwiched between a hadronic state and the vacuum defines the hadron wave
functions. When the LCQSR is used to calculate the coupling constant, the 
double Borel transformation is always invoked so that the excited states and 
the continuum contribution can be treated quite nicely. Moreover, the final 
sum rule depends only on the value of the hadron wave function at a 
specific point, which is much better known than the whole wave function \cite{bely95}. 
In the present case our sum rules invole with the pion wave function (PWF)
$\varphi_{\pi}(u_0 ={1\over 2})$. Note this parameter 
is universal in all processes at a given scale.
In this respect, $\varphi_{\pi}(u_0 ={1\over 2})$ is a fundamental 
quantity like the quark condensate. In \cite{bely-z} the value is 
estimated as $\varphi_{\pi}(u_0 ={1\over 2})=1.2\pm 0.3$ 
using the pion nucleon coupling constant and the 
phenomenological $\rho\omega\pi$ coupling constant as inputs. 
Recently the light cone sum rule for $g_{\pi NN} (q^2=0)$ \cite{bely-z} 
was reanalyzed \cite{zhu0}. The contribution 
from the gluon condensate $<g_s^2 G^2>$ and the quark gluon mixed condensate 
$\langle g_c {\bar q}\sigma \cdot G q\rangle$ is added. 
The uncertainty due to $\lambda_N$ is reduced in the numerical analysis 
with the help of the Ioffe's mass sum rule. A new value 
$\varphi_\pi (1/2)=1.5\pm 0.2$ \cite{zhu0} is obtained using the experimentally 
precisely known $g_{\pi NN}$ \cite{nijimegen}. 

The LCQSR has been widely used to derive the couplings of pions with heavy mesons 
in full QCD \cite{bely95}, in the limit of $m_Q\to \infty$ 
\cite{zhu1} and $1/m_Q$ correction \cite{zhu3}, the couplings 
of pions with heavy baryons \cite{zhu2}, the $\rho$-decay widths 
of excited heavy mesons \cite{zhu4} and various semileptonic decays of 
heavy mesons \cite{bagan98} etc.

Our paper is organized as follows: Section \ref{sec1} is an introduction.
We introduce the two point function for the $K^\ast K\pi$ vertex in section \ref{sec2}. 
The definitions of the PWFs are presented in section \ref{sec3}.
In the following section we present the LCQSR for the $K^\ast K\pi$ and 
$\rho\pi\pi$ couplings. A short summary is given in the last section.

\section{Two Point Correlation Function for the $K^\ast K\pi$ coupling}
\label{sec2}

The dominant decay mode is $K^\ast\to K\pi$ for $K^\ast$ and 
$\rho\to\pi\pi$ for $\rho$. The relevant decay amplitudes are
\begin{equation}
<K^{*0} (p) \pi^- (q) |K^- (p+q) >=-g_{K^* K\pi} q_\mu \epsilon^\mu
\end{equation}
\begin{equation}
<K^{*-} (p) \pi^0 (q) |K^- (p+q) >=-{g_{K^* K\pi}\over \sqrt{2}} q_\mu \epsilon^\mu
\end{equation}
\begin{equation}
<\rho^- (p) \pi^0 (q) |\pi^- (p+q) >=-g_{\rho\pi\pi} q_\mu \epsilon^\mu
\end{equation}
where $\epsilon_\mu$ is the polarization vector of the vector meson.
The resulting decay width reads 
\begin{equation}
\Gamma (K^*\rightarrow K\pi )=
\frac{ (g_{K^* K\pi})^2|q_{\pi}|^3 }{16\pi m_{K^*}^2}
\end{equation}
\begin{equation}
\Gamma (\rho\rightarrow \pi\pi )=
\frac{ (g_{\rho\pi\pi})^2|q_{\pi}|^3 }{24\pi m_{\rho}^2}
\end{equation}
With $\Gamma (K^*\rightarrow K\pi )=50.8$MeV and 
$\Gamma(\rho\rightarrow\pi\pi )=154$MeV \cite{PDG} one 
gets $g_{K^* K\pi}=9.08$ and $g_{\rho\pi\pi}=12.16$.  

To study these couplings, we start with the two-point correlation function:
\begin{equation}\label{eq2}
\Pi (p)=i\int d^4x e^{ipx}\langle \pi^- (q) | T [ {\bar d}(x)\gamma^\mu s(x), 
{\bar s}(0)\gamma^\alpha \gamma_5 u(0) ] |0\rangle \, .
\end{equation}
The instanton contributions may invalidate the usual 
sum rule techniques for the pseudo-scalar current \cite{Novikov}. We use 
the pseudo-vector currents for $K$ and $\pi$ throughout this work.

At the phenomenological level the eq.(\ref{eq2}) can be expressed as:
\begin{equation}\label{pole}
\Pi (p_1,p_2,q)  =   g_{K^\ast K\pi}
{  f_{K^\ast} m_{K^\ast} \over (p^2 - m_{K^\ast}^2)}
{f_K q_\beta p_2^\alpha \over (p_2^2 - m_K ^2) } 
(g^{\mu\beta}- {p^\mu p^\beta \over m^2_{K^\ast}})  +\cdots
\end{equation}
with $p_1 =p$, $p_2 = p+q$. 
The ellipse denotes the continuum and the off-diagonal transition contribution. 
The decay constants $f_K$ and $f_{K^\ast}$ are introduced as:
\begin{equation}
\langle K (p)|{\bar s}(0) \gamma_\mu \gamma_5 u(0)|0 \rangle =f_K p_\mu \, ,
\end{equation}
\begin{equation}
\langle K^\ast |{\bar d}(0) \gamma_\mu s(0)|0 \rangle =
m_{K^\ast}f_{K^\ast}\epsilon_\mu \, ,
\end{equation}
where $\epsilon_{\mu}$ is the polarization vector of the $K^\ast$ meson, 

After discarding the single pole terms in (\ref{pole}) which will always 
be eliminated through double Borel transformation later, 
the polarization operator can be expressed as 
\begin{equation}\label{pole1}
\Pi (p_1,p_2,q)  =   g_{K^\ast K\pi}
{  f_{K^\ast} m_{K^\ast} \over (p^2 - m_{K^\ast}^2)}
{f_K p_2^\alpha \over (p_2^2 - m_K ^2) } 
[q^\mu +{1\over 2} (1-{m_K^2 \over m^2_{K^\ast}})p^\mu]  \; .
\end{equation}
In the following we shall focus on the tensor structure $p^\alpha q^\mu$.

\section{The Formalism of LCQSR and Pion Wave Functions}
\label{sec3}
Neglecting the four particle component of the pion wave function, 	
the expression for $\Pi (p_1,p_2,q)$ 
with the tensor structure at the quark level reads,
\begin{equation}\label{quark}
\Pi (p_1,p_2,q) = -{\rm i} \int d^4x e^{ipx}
{\bf Tr} \{ \langle \pi^- (q) | u(0) {\bar d}(x)  |0\rangle 
\gamma^\mu iS_s (x) \gamma^\alpha \gamma_5 \} \; .
\end{equation}
where $iS_s(x)$ is the full strange quark propagator with both perturbative  
term and contribution from vacuum fields. 
\begin{eqnarray}\label{prop}\nonumber
iS_s(x)=\langle 0 | T [s(x), {\bar s}(0)] |0\rangle 
=-i\int {d^4k\over (2\pi)^4 }e^{-ikx} { {\hat k} +m_s \over (m_s^2 -k^2)}&\\ \nonumber
-ig_s\int {d^4k\over (2\pi)^4} e^{-ikx}\int_0^1 dv 
[{1\over 2}{ {\hat k} +m_s \over (m_s^2 -k^2)^2}
G^{\mu\nu}(vx)\sigma_{\mu\nu} +{1\over m_s^2-k^2}vx_\mu G^{\mu\nu}(vx)\gamma_\nu] &\\
-{\langle {\bar s} s\rangle  \over 12}
-{x^2 \over 192}\langle {\bar s}g_s \sigma\cdot G s\rangle 
+\cdots  & \; .
\end{eqnarray}
where we have introduced ${\hat k} \equiv k_\mu \gamma^\mu$, $m_s=150$MeV is the 
strange quark mass, $D_\mu =\partial_\mu -ig_s A_\mu$.

By the operator expansion on the light-cone
the matrix element of the nonlocal operators between the vacuum and 
pion state defines the two particle pion wave function.
Up to twist four the Dirac components of this wave function can be 
written as \cite{bely95}:
\begin{eqnarray}\label{phipi}
<\pi(q)| {\bar d} (x) \gamma_{\mu} \gamma_5 u(0) |0>&=&-i f_{\pi} q_{\mu} 
\int_0^1 du \; e^{iuqx} (\varphi_{\pi}(u) +x^2 g_1(u) + {\cal O}(x^4) ) 
\nonumber \\
&+& f_\pi \big( x_\mu - {x^2 q_\mu \over q x} \big) 
\int_0^1 du \; e^{iuqx}  g_2(u) \hskip 3 pt  , \label{ax} \\
<\pi(q)| {\bar d} (x) i \gamma_5 u(0) |0> &=& {f_{\pi} m_{\pi}^2 \over m_u+m_d} 
\int_0^1 du \; e^{iuqx} \varphi_P(u)  \hskip 3 pt ,
 \label{phip}  \\
<\pi(q)| {\bar d} (x) \sigma_{\mu \nu} \gamma_5 u(0) |0> &=&i(q_\mu x_\nu-q_\nu 
x_\mu)  {f_{\pi} m_{\pi}^2 \over 6 (m_u+m_d)} 
\int_0^1 du \; e^{iuqx} \varphi_\sigma(u)  \hskip 3 pt .
 \label{psigma}
\end{eqnarray}
\noindent 

\begin{eqnarray}
& &<\pi(q)| {\bar d} (x) \sigma_{\alpha \beta} \gamma_5 g_s 
G_{\mu \nu}(ux)u(0) |0>=
\nonumber \\ &&i f_{3 \pi}[(q_\mu q_\alpha g_{\nu \beta}-q_\nu q_\alpha g_{\mu \beta})
-(q_\mu q_\beta g_{\nu \alpha}-q_\nu q_\beta g_{\mu \alpha})]
\int {\cal D}\alpha_i \; 
\varphi_{3 \pi} (\alpha_i) e^{iqx(\alpha_1+v \alpha_3)} \;\;\; ,
\label{p3pi} 
\end{eqnarray}

\begin{eqnarray}
& &<\pi(q)| {\bar d} (x) \gamma_{\mu} \gamma_5 g_s 
G_{\alpha \beta}(vx)u(0) |0>=
\nonumber \\
&&f_{\pi} \Big[ q_{\beta} \Big( g_{\alpha \mu}-{x_{\alpha}q_{\mu} \over q \cdot 
x} \Big) -q_{\alpha} \Big( g_{\beta \mu}-{x_{\beta}q_{\mu} \over q \cdot x} 
\Big) \Big] \int {\cal{D}} \alpha_i \varphi_{\bot}(\alpha_i) 
e^{iqx(\alpha_1 +v \alpha_3)}\nonumber \\
&&+f_{\pi} {q_{\mu} \over q \cdot x } (q_{\alpha} x_{\beta}-q_{\beta} 
x_{\alpha}) \int {\cal{D}} \alpha_i \varphi_{\|} (\alpha_i) 
e^{iqx(\alpha_1 +v \alpha_3)} \hskip 3 pt  \label{gi} 
\end{eqnarray}
\noindent and
\begin{eqnarray}
& &<\pi(q)| {\bar d} (x) \gamma_{\mu}  g_s \tilde G_{\alpha \beta}(vx)u(0) |0>=
\nonumber \\
&&i f_{\pi} 
\Big[ q_{\beta} \Big( g_{\alpha \mu}-{x_{\alpha}q_{\mu} \over q \cdot 
x} \Big) -q_{\alpha} \Big( g_{\beta \mu}-{x_{\beta}q_{\mu} \over q \cdot x} 
\Big) \Big] \int {\cal{D}} \alpha_i \tilde \varphi_{\bot}(\alpha_i) 
e^{iqx(\alpha_1 +v \alpha_3)}\nonumber \\
&&+i f_{\pi} {q_{\mu} \over q \cdot x } (q_{\alpha} x_{\beta}-q_{\beta} 
x_{\alpha}) \int {\cal{D}} \alpha_i \tilde \varphi_{\|} (\alpha_i) 
e^{iqx(\alpha_1 +v \alpha_3)} \hskip 3 pt . \label{git} 
\end{eqnarray}
\noindent 
The operator $\tilde G_{\alpha \beta}$  is the dual of $G_{\alpha \beta}$:
$\tilde G_{\alpha \beta}= {1\over 2} \epsilon_{\alpha \beta \delta \rho} 
G^{\delta \rho} $; ${\cal{D}} \alpha_i$ is defined as 
${\cal{D}} \alpha_i =d \alpha_1 
d \alpha_2 d \alpha_3 \delta(1-\alpha_1 -\alpha_2 
-\alpha_3)$. 
Due to the choice of the
gauge  $x^\mu A_\mu(x) =0$, the path-ordered gauge factor
$P \exp\big(i g_s \int_0^1 du x^\mu A_\mu(u x) \big)$ has been omitted.
The coefficient in front of the r.h.s. of 
eqs. (\ref{phip}), (\ref{psigma})
can be written in terms of the light quark condensate
$<{\bar u} u>$ using the PCAC relation: 
$\displaystyle \mu_{\pi}= {m_\pi^2 \over m_u+m_d}
=-{2 \over f^2_\pi} <{\bar u} u>$. 

The PWF $\varphi_{\pi}(u)$ is associated with the leading twist two 
operator, $g_1(u)$ and $g_2(u)$ correspond to twist four operators, 
and $\varphi_P(u)$ and $\varphi_\sigma (u)$ to twist three ones. 
The function $\varphi_{3 \pi}$ is of twist three, while all the 
PWFs appearing in eqs.(\ref{gi}), (\ref{git}) are of twist four.
The PWFs $\varphi (x_i,\mu)$ ($\mu$ is the renormalization point) 
describe the distribution in longitudinal momenta inside the pion, the 
parameters $x_i$ ($\sum_i x_i=1$) 
representing the fractions of the longitudinal momentum carried 
by the quark, the antiquark and gluon.

The wave function normalizations immediately follow from the definitions
(\ref{phipi})-(\ref{git}):
$\int_0^1 du \; \varphi_\pi(u)=\int_0^1 du \; \varphi_\sigma(u)=1$,
$\int_0^1 du \; g_1(u)={\delta^2/12}$,
$\int {\cal D} \alpha_i \varphi_\bot(\alpha_i)=
\int {\cal D} \alpha_i \varphi_{\|}(\alpha_i)=0$,
$\int {\cal D} \alpha_i \tilde \varphi_\bot(\alpha_i)=-
\int {\cal D} \alpha_i \tilde \varphi_{\|}(\alpha_i)={\delta^2/3}$,
with the parameter $\delta$ defined by the matrix element: 
$<\pi(q)| {\bar d} g_s \tilde G_{\alpha \mu} \gamma^\alpha u |0>=
i \delta^2 f_\pi q_\mu$.

\section{The LCQSR for the $K^\ast K \pi$ coupling}
\label{sec4}
Expressing (\ref{quark}) with the strange quark propagator and 
keeping only the tensor structure $p^\alpha q^\mu$, we arrive at:
\begin{eqnarray}\label{coordinate}\nonumber
\Pi (p_1, p_2, q) = 
-if_\pi \int_0^1 {du\over m_s^2 -(p+uq)^2} \{
\varphi_\pi (u) &\\ \nonumber
+{ {m_s\over 3}\mu_\pi \varphi_\sigma (u) +4ug_2(u) -4g_1(u)-4G_2(u)
\over  m_s^2 -(p+uq)^2} 
-{8m_s^2 [g_1(u)+G_2(u)]\over [m_s^2 -(p+uq)^2]^2 }
+\cdots \} \;,
\end{eqnarray}
where $\mu_{\pi}=1.65$GeV, $f_{\pi}=132$MeV, 
$G_2 (u)=-\int_0^{u} g_2(u)du $, which arises from  integration by parts 
to absorb the factor $1/(q\cdot x)$,  
\begin{equation}\label{integration}
\int_0^1 {e^{-iu q\cdot x}\over  q\cdot x }g_2 (u)  du =
-i\int_0^1  e^{-iu q\cdot x} G_2 (u)  du -
G_2 (u) e^{-iu q\cdot x}|_0^1 \; ,
\end{equation}
Note the second term in (\ref{integration}) vanishes due to 
$G_2 (u_0)=0$ at end points $u_0 =0, 1$.

Making double Borel transformation with the variables $p_1^2$ and $p_2^2$
the single-pole terms in (\ref{pole}) are eliminated. The formula reads:
\begin{equation}\label{double}
{{\cal  B}_1}^{M_1^2}_{p_1^2} {{\cal  B}_2}^{M_2^2}_{p_2^2} 
{\Gamma (n)\over [ m^2 -(1-u)p_1^2-up^2_2]^n }=
(M^2)^{2-n} e^{-{m^2\over M^2}} \delta (u-u_0 ) \; .
\end{equation}

Subtracting the continuum contribution which is modeled by the 
dispersion integral in the region $s_1 , s_2\ge s_0$, we arrive at:
\begin{eqnarray}\label{quark0}\nonumber
f_K f_{K^\ast} m_{K^\ast} g_{K^\ast K\pi} e^{-{m_K^2+m_{K^\ast}^2\over 2M^2}}=
f_\pi e^{-{u_0(1-u_0)q^2+m_s^2\over M^2}} \{
M^2 (1-e^{-{s_1\over M^2}})\varphi_\pi (u_0) &\\ 
+{m_s\over 3}\mu_\pi \varphi_\sigma (u_0)
+4u_0g_2(u_0)
-4g_1(u_0)-4G_2(u_0)-{4m_s^2\over M^2}[g_1(u_0)+G_2(u_0)]
\}&\; ,
\end{eqnarray}
where $s_1$ is the continuum threshold,
$u_0={M^2_1 \over M^2_1 + M^2_2}$, 
$M^2\equiv {M^2_1M^2_2\over M^2_1+M^2_2}$, 
$M^2_1$, $M^2_2$ are the Borel parameters. 
Note all the PWFs involved with the vacuum gluon field do not contribute 
to the tensor structure $p^\alpha q^\beta$, which greatly simplifies the
analysis of the sum rule.

Similarly we can obtain the LCQSR for $\rho\pi\pi$ coupling. In our calculation
we take the up and down quark current mass to be zero. Then we arrive at the
simple sum rule:
\begin{eqnarray}\label{quark1}\nonumber
f_\rho  m_\rho g_{\rho\pi\pi} e^{-{m_\pi^2+m_{\rho}^2\over 2M^2}}=  
\sqrt{2} e^{-{u_0(1-u_0)q^2+m_s^2\over M^2}} \{
M^2 (1-e^{-{s_2\over M^2}})\varphi_\pi (u_0) &\\ 
+4u_0g_2(u_0)-4g_1(u_0)-4G_2(u_0)\}&\; .
\end{eqnarray}

The sum rule (\ref{quark0}) and (\ref{quark1}) appears asymmetric with the 
different masses for $K$ and $K^\ast$ mesons at first sight. 
However, if a sum rule holds well, there should exist a certain interval called 
the working region of the Borel parameter, $M_i^2 < M_B^2 < M^2_f$. 
Within this region the sum rule is insensitive to $M_B^2$ and stays 
reasonably stable with the variation of $M_B$. 
In other words, every point of $M_B$ in the above interval is equally
good for the analysis of the sum rule.  
So long as the working regions for $M_1^2$ and $M^2_2$ have some 
overlapping region, which does occur in our case, 
we can choose a common value in the overlapping region for 
both $M^2_1$ and $M^2_2$. From the above argument we know such a choice 
will not alter significantly the final result of the sum rule, i.e., the  
choice $M^2_1 =M^2_2$ is allowed in the analysis of the LCQSR.
Moreover, the symmetric choice of $M^2_1 =M^2_2$ will enable the clean subtraction
of the continuum contribution, which is crucial for the numerical 
analysis of the sum rules. In contrast, the asymmetric choice
will lead to the continuum subtraction extremely difficult in \cite{bely95}, 
which is the operational and technical motivation for the choice of $u_0=1/2$. 
We shall work in the physical limit $q^2=m_\pi^2 \to 0$ in (\ref{quark0}).

The resulting sum rule depends on the PWFs and the integrals of them 
at the point $u_0={1\over 2}$. We adopt $\varphi_\pi(u_0)=1.5\pm 0.2$ \cite{zhu0}. 
For the other PWFs we use the results given in \cite{bely95},  
$\varphi_\sigma(u_0)=1.47$,  
$g_1(u_0)=0.022 $GeV$^2$, $g_2 (u_0) =0$ and 
$G_2(u_0)=0.02$GeV$^2$ at $u_0 ={1\over 2}$ at the scale $\mu =1$GeV. 

The overlap amplitudes $f_{K^\ast}$ and $f_\rho$ can be determined in 
a self-consistent manner making use of the corresponding mass sum rules. 
For example, for the $\rho$ meson \cite{SVZ,RRY}, we have:
\begin{equation}
\label{rho}
m_\rho^2 f_{\rho}^2 =\frac{1}{\pi^2} e^{\frac{m_{\rho}^2}{M^2_B}} \{ \frac{1}{4}
 (1+\frac{\alpha_s}{\pi}) M^4_B E_1 -\frac{b}{48} +
\frac{\alpha_s}{\pi}\frac{14}{81} a^2_q  \frac{1}{M^2_B} \},
\end{equation}
where $E_1\equiv 1-(1+\frac{s_0}{M^2_B})e^{-\frac{s_0}{M^2_B}}$ (with the 
continuum threshold $s_0=1.5$GeV$^2$) is the factor used to subtract the 
continuum contribution. 
Numerically, $f_\rho =(0.18 \pm 0.02)$GeV, $f_{K^\ast}=(0.21 \pm 0.02)$GeV,
$f_K=(0.15\pm 0.02)$GeV \cite{SVZ,RRY,PDG,rev}.

Note that we have used the pseudo-vector interpolating current, 
which couples strongly to both pseudo-scalar mesons 
$\pi$, $K$ and pseudo-vector mesons $a_1 (1260)$, $K_1 (1270)$. $a_1 (1260)$ 
is a broad resonance with a full width of $\sim 200$MeV. In order to 
eliminate the contamination from $a_1(1260)$ and $K_1(1270)$, we choose 
the continuum threshold parameter to be 
$s_2 \le (m_{a_1}-{\Gamma_{a_1}\over 2})^2 \sim (1.2\pm 0.1)$GeV$^2$ 
and $s_1 \le (m_{K_1}-{\Gamma_{K_1}\over 2})^2 \sim (1.5\pm 0.1)$GeV$^2$ 
These values are consistent 
with the continuum threshold for the sum rules of pseudo-scalar mesons. But 
they are slightly smaller than the continuum threshold for vector mesons.

The final sum rules are stable with reasonable variation of 
the Borel mass around $m^2_{K^\ast}$ and $m_\rho^2$ respectively
after the exponential factor is moved to the left hand side as can be seen
from Fig. 1 and Fig. 2. The terms with the strange quark mass $m_s$ 
contribute about $10\%$ to the whole sum rule (\ref{quark0}).
Our final result is $g_{K^\ast K\pi}=(8.7\pm 0.5)$ and 
$g_{\rho\pi\pi}=(11.5\pm 0.8)$ with the central values of $f_\rho, f_K,
f_{K^*}$, which agrees very well with the value extracted from 
the experimental data
$g_{K^\ast K\pi}=9.08$ and $g_{\rho\pi\pi}=12.16$ \cite{PDG}.

Although the coupling of $K^*\rho K$ may be sizeable, the $\rho K$ intermediate
states contribute to the continuum only because $m_\rho +m_K =1.27\mbox{GeV}
> \sqrt{s_1}$. In our approach we have invoked quark-hadron duality 
and modeled the spectral density with $s>s_1$ at the hadronic side
with the free parton-like one. So the $\rho K$ contribution is 
subtracted away. On the other hand, the intermediate state $3\pi K$ does lie 
around $m_{K^*}$ and below the continuum threshold $\sqrt{s_1}$. 
But due to the strong suppression from the four-body phase space integral,
its contribution is negligible. In other words, the $K^*$ pole term dominates
the possible intermediate states with same quantum numbers 
below the continuum threshold $s_1$ while those intermediate states
above $s_1$ is subtracted away using the quark-hadron duality assumption,
which is the corner stone of the QCD sum rules approach.

In summary we have calculated $K^\ast K\pi$ and $\rho\pi\pi$ couplings using
the light cone QCD sum rules. These couplings are related to the values of 
PWFs at the point $u_0={1\over 2}$, which are universal in all processes. 
Our results are in good agreement with the experimental data. 

\vspace{0.8cm} {\it Acknowledgments:\/} 
This work was supported by the National Natural Science Foundation of China
and the Postdoctoral Science Foundation of China.

{\bf Figure Captions}
\vspace{2ex}
\begin{center}
\begin{minipage}{130mm}
{\sf FIG 1.} \small{The sum rule for $g_{\rho\pi\pi}$ as a functions of the Borel 
parameter $M^2$.  From bottom to top the curves correspond to $s_2 =
1.1, 1.2, 1.3$GeV$^2$.
}
\end{minipage}
\end{center}
\begin{center}
\begin{minipage}{130mm}
{\sf FIG 2.} \small{The sum rule for $g_{K^\ast K\pi}$ as a functions of the Borel 
parameter $M^2$.  From bottom to top the curves correspond to $s_1 =
1.4, 1.5, 1.6$GeV$^2$. 
}
\end{minipage}
\end{center}

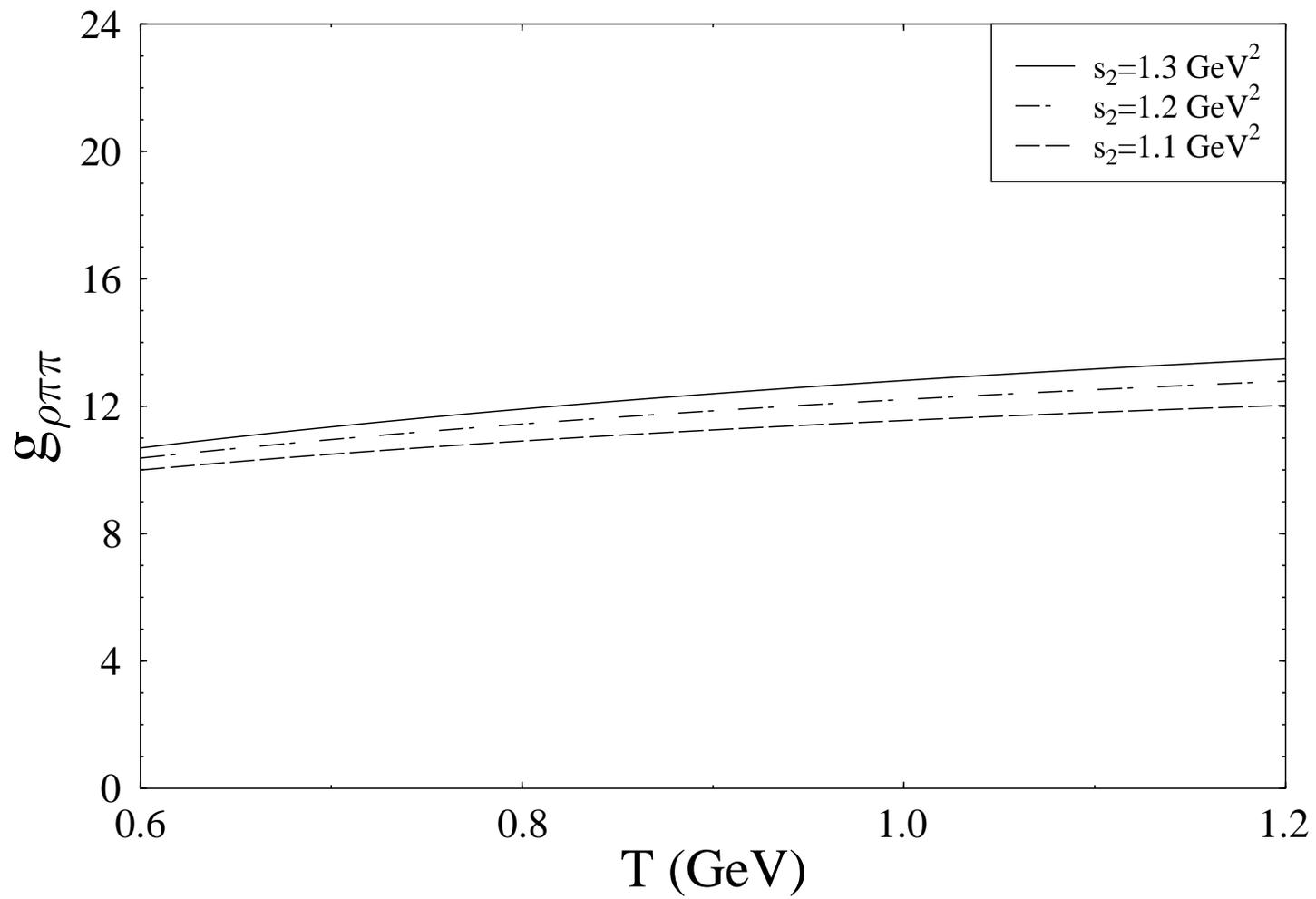
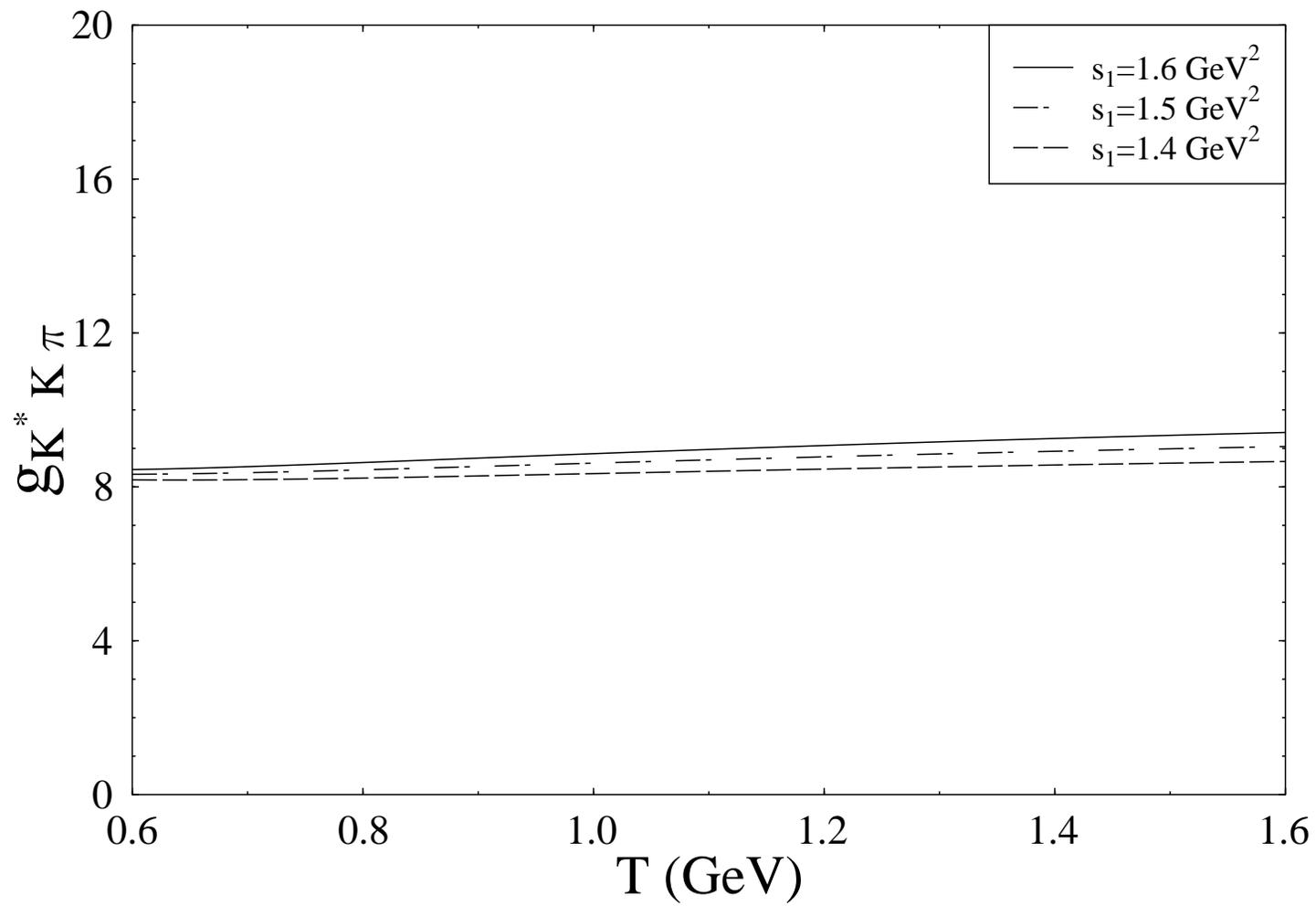
\end{document}